\documentclass[12pt]{article}
\usepackage{graphicx}
\usepackage{amssymb}
\usepackage{amsfonts}

\catcode`@=11
\def\secteqno{\@addtoreset{equation}{section}%
\def\theequation{\thesection.\arabic{equation}}}
\catcode`@=12
\secteqno
\newcommand{\fr}[1]{\mathfrak{#1}}

\newcommand{\be}{\begin{equation}}
\newcommand{\ee}{\end{equation}}
\newcommand{\bea}{\begin{eqnarray}}
\newcommand{\eea}{\end{eqnarray}}

\newcommand{\bref}[1]{(\ref{#1})}
\newcommand{\nn}{\nonumber}	   
\newcommand{\T}{\frac{1}{2\pi\alpha'}}

\mathcode`\*="702A                  
\catcode163=13 \def\itm{\relax\ifmmode\to\else\itemize\fi}
\newcommand{\beq}{\begin{equation}}
\newcommand{\eeq}{\end{equation}}
\newcommand{\beqa}{\begin{eqnarray}}
\newcommand{\eeqa}{\end{eqnarray}}
\begin{document}
\thispagestyle{empty}
\hfill March 13, 2006

\hfill KEK-TH-1078

\vskip 20mm
\begin{center}
{\Large\bf Nonlocal charges of T-dual strings}
\vskip 6mm
\medskip
\vskip 10mm
{\large Machiko\ Hatsuda$^{\ast\dagger a}$ and Shun'ya\ Mizoguchi$^{\ast b}$}

\parskip .15in
{\it $^\ast$Theory Division,\ High Energy Accelerator Research Organization (KEK),\\
\ Tsukuba,\ Ibaraki,\ 305-0801, Japan} \\{\it $^\dagger$Urawa University, Saitama \ 336-0974, Japan}\\
\vspace*{1cm}
$^{a}$mhatsuda@post.kek.jp
~~~~
 $^{b}$mizoguch@post.kek.jp
\end{center}

\parskip .25in

\medskip
\vskip 10mm
\begin{abstract}
We obtain sets of infinite number of conserved nonlocal charges of strings 
in a flat space and pp-wave backgrounds,
and  compare them before and after T-duality transformation.
In the flat background 
the set of nonlocal charges is the same before and after the
T-duality transformation with interchanging odd and even-order charges.
In the IIB pp-wave background
an infinite  number of nonlocal charges
are independent,
contrast to that in a flat background 
only the zero-th and first 
order charges are independent.
In the IIA pp-wave background,
which is the T-dualized 
compactified IIB pp-wave background,
the zero-th order charges are included 
as a part of the set of nonlocal charges in the IIB 
background.
To make this correspondence complete
a variable conjugate to the winding number 
is introduced as a Lagrange multiplier in the IIB action
$\grave{\rm a}$ la Buscher's transformation.
\end{abstract} 

\setcounter{page}{1}
\parskip=7pt
\newpage
\section{ Introduction}

Integrability is one of the most important guiding 
principles to find correspondences
beyond the BMN limit \cite{BMN} of  AdS/CFT correspondence.
After generalizations of  BMN's work \cite{GKP2,FT} 
 the Bethe ansatz was brought to use in this problem \cite{MZ}, 
and integrable spin chain models have been shown to 
describe
semiclassical spinning strings 
\cite{Frolov:2003xy} as well as
superconformal Yang-Mills theories
 \cite {Beisert:2003tq}.
 Those techniques
are used to examine the correspondence between
the string energy spectrum 
and the dilatation operator 
 \cite{Arutyunov:2003uj,Beisert:2003yb}
 (see \cite{Zarembo:2004hp} for review).
For an integrable system
conserved quantities are not only
 the above energy operator but also
an infinite set of conserved nonlocal charges
\cite{LP}.
The existence of an infinite set of conserved nonlocal charges 
 implies the absence of particle production and the factorization of the S-matrix
\cite{LP,Lipatov}.

Brezin, Itzykson, Zinn-Justin and Zuber (BIZZ) 
gave a simple derivation of these nonlocal charges
\cite{BIZZ}: Assume that one has a set  of conserved and ``flat" currents $J_\alpha$, 
$\partial_\alpha J^\alpha=0$ and
 $\epsilon^{\alpha\beta}
(\partial_\alpha J_\beta-J_\alpha J_\beta)=0$.
Then an infinite set of nonlocal currents ${\cal J}_{[N]\alpha}$ are constructed
inductively.
The zero-th order current is set to be the Noether current ${\cal J}_{[0]\alpha}=J_\alpha$.
The first order current, ${\cal J}_{[1]\alpha}$, is determined by
 the ``flatness" condition as
$\epsilon^{\alpha\beta}
(\partial_\alpha J_\beta-J_\alpha J_\beta)=\partial_\alpha {\cal J}_{[1]}^\alpha=
0$ where
the nonabelian structure of $J_\alpha$ 
gives rise to the nonlocality of ${\cal J}_{[N]\alpha}$.
The ``flatness" condition is a ``dual" equation
of the two-dimensional worldsheet which is closely related to 
the
 Buscher T-duality
\cite{Buscher:1987sk}.
In this paper we examine whether 
the two-dimensional dual formulation required 
in the nonlocal charges shares
any concept with the two-dimensional transformation of
T-duality.

Bena, Roiban and Polchinski showed
the existence of  an infinite set of conserved 
nonlocal charges of a superstring in the AdS$_5\times$S$^5$ background 
\cite{BPR},
while the one for the bosonic part was shown in \cite{Mandal:2002fs}
and the one in manifestly $\kappa$-invariant way was shown in 
\cite{HY}.
Additional discussion of nonlocal charges 
of strings in the AdS background
can be found in refs. 
\cite{Vallilo:2003nx,Berkovits:2004jw,
Hou:2004ru,Wolf:2004hp,Das:2004hy,Alday:2005gi,Young:2005jv,{Miller:2006bu}}.
For an integrable model there exist a spectral parameter
which relates
 an infinite  set of ``local" charges and
an infinite set of nonlocal charges.
 Recently it has been shown that 
an infinite number of conserved ``local" charges
for a superstring in the AdS$_5\times$S$^5$ background
are combinations of
the Virasoro constraint and the $\kappa$ symmetry constraints
\cite{Hatsuda:2005te} which govern
 the target space field equations.
So it is natural to expect
 that the nonlocal charges
reflect the spacetime isometry and 
stringy symmetry of the target space.
In order to figure out  physical meanings of 
nonlocal charges
we compute  nonlocal charges explicitly.
In this paper a flat space background 
and the IIB and IIA pp-wave backgrounds are examined
so the nonlocal integration
can be performed in terms of oscillators.

The existence of an infinite set of 
nonlocal charges in the pp-wave background 
was shown by Alday \cite{Alday:2003zb}. 
We will further develop the work by  
(1)  computing higher order nonlocal charges with respect to complete
set of isometry generators,
and
(2)  computing nonlocal charges of a string in the pp-wave
background before and after T-dual transformation.
The IIB  pp-wave background  is compactified 
and T-duality transformation is performed as 
\cite{Michelson:2002wa}, then 
the string in the IIA pp-wave background
is quantized \cite{Alishahiha:2002nf,Mizoguchi:2003be}.
For the IIA background we will focus on the Noether charges 
which are the zero-th order
charges,
and we will examine whether they are related to the nonlocal charges of the 
IIB background.

We also clarify the procedure of
constructing nonlocal charges 
of strings in a flat and 
the pp-wave background which are described by
 the inhomogeneous O($n$) nonlinear sigma model.
The BIZZ procedure \cite{BIZZ} 
was originally applied to a homogeneous O($n$)
non-linear sigma model;
the nonlocal charges are systematically
constructed 
from a conserved and ``flat" current.
This current  is a O($n$) left-invariant current 
and satisfies 
the Maurer-Cartan (M.C.) equation for the o($n$) algebra
which is the ``flatness" condition. 
For a string in a flat  background
the geometry may be regarded as a coset sigma model 
of the Poincar$\acute{\rm e}$ group over the Lorentz group
and the string coordinate is the coset parameter.
But the left-invariant current $J_\alpha=g^{-1}\partial_\alpha g$
of a coset element $g(X^\mu)={\rm exp}(iX^\mu P_\mu)$ 
neither has a Lorentz generator component
nor satisfies the M.C. equation 
for the Poincar$\acute{\rm e}$ algebra,
because the coset algebra is abelian, $\left[P_\mu,P_\nu\right]=0$.     
So the BIZZ procedure is not applicable naively to 
a  string in a flat space.
On the other hand 
the flat background is  obtained by the 
In$\ddot{\rm o}$n$\ddot{\rm u}$-Wigner (IW)
contraction of an AdS space
represented by the coset of 
the homogeneous SO($n,2$) group over the SO($n,1$),
then one-form currents obtained by the IW contraction
 satisfy the Poincar$\acute{\rm e}$  M.C. equation.
For the AdS space before the IW contraction 
 the  conserved  and ``flat" current
is nothing but the Noether current of the action.
From above two facts 
we begin with Noether currents  as the 
conserved currents
for a flat and pp-wave spaces, 
and we will examine whether they satisfy the ``flatness" condition 
corresponding to the background isometries.
Using these conserved and ``flat" currents
we will construct an infinite set of conserved nonlocal currents
and charges in a flat and pp-wave backgrounds.

We also point out how to define the integration path for
 conserved charges of a closed string 
 with winding modes.
 For a conserved current $\partial_\alpha J^\alpha=0$
 one utilizes a dual potential $\chi$ such that 
 $J_\alpha=\epsilon_{\alpha\beta}\partial^\beta \chi$,
but this $\chi$ is a single-valued function only on a simply
connected region. 
Therefore we define the potential $\chi$,
and hence the nonlocal current, 
on a semi-infinite strip cut open
the cylinderical worldsheet. 
 The conserved charges are given by integration along the boundary path
of the semi-infinite strip.

The organization of the paper is as follows. 
In section 2 the current conservation  and the ``flatness"
condition for a string in a flat background are
examined and 
a general argument of the Noether current for inhomogeneous backgrounds  
is presented. By the
BIZZ procedure  
we construct
nonlocal charges and compare them before and after  
the T-duality transformation.
In section 3 we compute nonlocal charges of a string in
the type IIB pp-wave background and write down 
concrete expressions for zero-th, first and second order.
In section 4 T-duality transformation on the Michelson's cycle
is performed to obtain the type IIA pp-wave background,
and nonlocal charges before and after the T-duality transformation are computed.
From the completeness of this correspondence,
a conjugate coordinate to the winding mode  is introduced
by adding the Wess-Zumino term in the type IIB action
which corresponds to the Buscher T-duality transformation. 

\section{ Flat background}

The action of a string in a flat background has not only the translational symmetry but also the Lorentz symmetry.
Although the translational group is abelian,
considering the whole Poincar$\acute{\rm e}$ group gives 
nontrivial structure of the nonlocal charges
especially under T-duality.
At first we will clarify the Noether current for 
inhomogeneous SO($n$)
cosets
and examine the procedure of constructing nonlocal charges.
Then we will compute nonlocal charges and compare them 
under T-duality.

\subsection{Noether charges}
The action for a string in a flat space is given by
\bea
S&=&-\frac{1}{4\pi\alpha'}\int~d^2\sigma~\sqrt{-h}h^{\alpha\beta}\partial_\alpha X^\mu\partial_\beta X^\nu \eta_{\mu\nu}
\eea
where $h_{\alpha\beta}$ is the worldsheet metric
and the conformal gauge $\sqrt{-h}h^{\alpha\beta}=\eta^{\alpha\beta}$ 
is chosen from now on.
A closed string is quantized as
\bea
&X^\mu=x^\mu+\alpha' p^\mu\tau+
\sqrt{\frac{\alpha'}{2}}\displaystyle\sum_{n\neq 0}
\frac{i}{n}\left(
\alpha_n^\mu e^{-in(\tau+\sigma)}+
\tilde{\alpha}_n^\mu e^{-in(\tau-\sigma)}
\right)~&\nn\\&
\left[x^\mu,p^\nu\right]=i\eta^{\mu\nu}~~,~~
\left[\alpha_m^\mu,\alpha_n^\nu\right]=m\delta_{m,-n}\eta^{\mu\nu}
=\left[\tilde{\alpha}_m^\mu,\tilde{\alpha}_n^\nu\right]&
~~~.
\eea

Noether currents under the translation and the Lorentz rotation are
\bea
&P^\mu_\alpha=\T \partial_\alpha X^\mu~~,~~M^{\mu\nu}_\alpha=\T X^{[\mu}\partial_\alpha X^{\nu]}
&\label{Noethercurrentsflat}
\eea
satisfying the conservation law
\bea
&\partial^\alpha P^{\mu}_\alpha=0=\partial^\alpha M^{\mu\nu}_\alpha~~~.&\label{cons}
\eea
Noether charges are given by
\bea
\hat{P}^\mu&=&\displaystyle\int_0^{2\pi}d\sigma~P^\mu_\tau =p^\mu~\nn\\~
\hat{M}^{\mu\nu}&=&\displaystyle\int_0^{2\pi}d\sigma~
M^{\mu\nu}_\tau =x^{[\mu}p^{\nu]}
+\frac{i}{2}\displaystyle\sum_{n\neq 0}\frac{1}{n}\left(
\alpha_n^{[\mu}\alpha_{-n}^{\nu]}
+\tilde{\alpha}_n^{[\mu}\tilde{\alpha}_{-n}^{\nu]}
\right)\label{flatPM0}
\eea
satisfying the Poincar$\acute{\rm e}$ algebra
\bea
\left[\hat{M}^{\mu\nu},\hat{M}^{\rho\lambda}\right]=i\eta^{[\mu\mid[\lambda}\hat{M}^{\rho]\mid\nu]}~~,~~
\left[\hat{M}^{\mu\nu}, \hat{P}^\rho\right]
=i\eta^{\rho[\mu}\hat{P}^{\nu]}~~.\label{PoiPoi}
\eea
The structure constant of the Poincar$\acute{\rm e}$ algebra in \bref{PoiPoi}
is denoted by $f_{ab}^{c}$ in $\left[T_a,T_b\right]=f_{ab}^{c}T_c$
with ${T}_a=(\hat{P}^\mu,~\hat{M}^{\mu\nu})$.

Next let us introduce the one forms by multiplying the one form basis as 
\bea
J^\mu=d\sigma^\alpha P^\mu_\alpha~~,~~
J^{\mu\nu}=d\sigma^\alpha M^{\mu\nu}_\alpha~~~,\label{1forms}
\eea
then we act the exterior derivative on them
\bea
dJ^\mu=0~~,~~
dJ^{\mu\nu}=4\pi\alpha'J^{\mu}\wedge J^{\nu}~~~\label{MCflat}.
\eea
In addition to the equations  \bref{MCflat}
the consistency $d dJ^A=0$ allows to regard 
the Noether currents \bref{Noethercurrentsflat} or \bref{1forms} 
as currents on a group manifold.
If we read off the ``structure constant" from 
\bref{MCflat} as
\bea
dJ^C=\frac{1}{2}F_{AB}^C J^A\wedge J^B~~,~~J^A=(J^\mu,~J^{\mu\nu})~~~,\label{dJJJ}
\eea 
then this structure constant $F_{AB}^C$ is different from the one 
of the Poincar$\acute{\rm e}$ algebra $f_{ab}^c$. 
This is  common for the inhomogeneous spaces such as a 
flat space and pp-wave spaces,
unlike AdS spaces.
At first we explain this feature and then we will use this
 structure constant $F_{AB}^C$
to obtain the nonlocal charges.

\par
\vskip 6mm


\subsection{Noether currents for inhomogeneous SO($n$) cosets}

Consider a $G/H$ coset sigma model with identification 
$g(x)\sim g(x)h(x)$, $g(x)\in G$, $h(x)\in H$. Let ${\cal G}$ and ${\cal H}$ be 
the Lie algebra  of  $G$ and $H$, and let ${\cal G}={\cal H}\oplus {\cal K}$. 
We assume that the coset is a symmetric space, that is, 
there exist a `parity' transformation $\theta$ such that 
\beqa
\theta({\cal H})~=~+{\cal H}~~,~~\theta({\cal K})~=~-{\cal K}~~.
\eeqa
As in \cite{BPR}, define 
\beqa
J_\alpha~=~g^{-1}\partial_\alpha g~~,~~
J_\alpha~=~H_\alpha+ K_\alpha~~
\eeqa
with $H_\alpha\in {\cal H}$, $K_\alpha \in {\cal K}$. 
The Lagrangian is given by 
\beqa
{\cal L}&=&\frac14 \mbox{Tr}\partial_\alpha {\cal M} \partial^\alpha {\cal M},~~~
{\cal M}=g\cdot \theta(g^{-1})\nonumber\\
&=&\mbox{Tr}K_\alpha K^\alpha. 
\eeqa
If the trace is non-degenerate on ${\cal K}$, one can 
derive the equation of motion  
\beqa
\partial_\alpha K^\alpha +{[}H_\alpha, K^\alpha{]}=0~~~.
\eeqa
Then it follows that $k_\alpha\equiv gK_\alpha g^{-1}$ is a conserved 
current:
\beqa
\partial_\alpha k^\alpha =0~~~.
\eeqa
In general $k_\alpha$ 
is invariant under the local gauge transformation, $g(x)\to g(x)h(x)$,
and is identified as the 
Noether current of the coset model.
The Noether current $k_\alpha$
has both ${\cal H}$ and ${\cal K}$ components unlike 
$K_\alpha$ which has only ${\cal K}$ components.
 However, if ${\cal K}$ is abelian, 
$k_\alpha$ belongs to ${\cal K}$ and has no ${\cal H}$ components.
 Therefore, 
in this case $k_\alpha$ constructed above does not exhaust all the Noether 
currents of the model. 
In our case the bosonic string action in the flat space is regarded 
as the coset sigma model of the Poincar$\acute{\rm e}$ group over the Lorentz group.  
Since the translation group is abelian, the conserved currents constructed 
as  $k_\alpha$ only contain $\partial_\alpha X^\mu$, the Noether currents for the 
translation,  but not those for the Lorentz rotations.

For a supersymmetric system the coset is not a symmetric space anymore
because of the superalgebra $\left\{Q,Q\right\}=P$.
The `parity' is modified to four-fold grading 
\cite{Berkovits:1999zq,Frolov:2006cc},
and the conserved current is not simply $gK_\alpha g^{-1}$ but
with the $\kappa$-symmetric modification \cite{HY}.

Let us consider a flat space as a coset ISO($D,1$)/SO($D-1,1$)  
obtained by the IW  contraction of a coset 
SO($D,1$)/SO($D-1,1$). 
We begin with
a coset element $g\in$
$G/H$ with $G$=SO($D,1$) and $H$=SO($D-1,1$).
The Noether current has both ${\cal H}$ and ${\cal K}$
components as $k=k_{\cal H}+k_{\cal K}$
with $k_{\cal H}\in {\cal H}  $
and $k_{\cal K}\in {\cal K}$.
They satisfy the following M.C. equation 
 \bea
dk_{\cal K}~=~k_{\cal H}\wedge k_{\cal K} ~~,~~ dk_{\cal H}~=~k_{\cal H}\wedge k_{\cal H}+k_{\cal K}\wedge k_{\cal K}~~~.
 \eea  
The IW contraction into  the Poincar$\acute{\rm e}$ algebra 
reduces to the following M.C. equations
\bea
k_{\cal K}\to \Omega k_{\cal K}~~,~~ k_{\cal H}\to k_{\cal H}~~{\rm and }~~\Omega\to 0~~~,~~{\rm then}~~
dk_{\cal K}~=~k_{\cal H}\wedge k_{\cal K}~~,~~dk_{\cal H}~=~k_{\cal H}\wedge k_{\cal H} ~~~.
\eea
These are not the equations \bref{MCflat}.
This contraction for the current is realized
if a coset element contains 
auxiliary coordinates $Y^{\mu\nu}$ in addition to the 
coset parameters $X^\mu$:
i.e. a coset element and the one-form currents are functions of
$X^\mu$ and $Y^{\mu\nu}$.
If they are rescaled as $X^\mu \to \Omega X^\mu$ and $Y^{\mu\nu}\to Y^{\mu\nu}$ 
then the currents can be rescaled as $k_{\cal K}\to\Omega k_{\cal K}$ and $k_{\cal H}\to k_{\cal H}$.
But if a coset element contains only $X^\mu$ then
rescaling $X^\mu\to\Omega X^\mu$ forces the currents
the following rescaling by the parity requirement
\bea
k_{\cal K}\to \Omega k_{\cal K}~~,~~ k_{\cal H}\to \Omega^2 k_{\cal H}~~{\rm and }~~\Omega\to 0~~,~~
{\rm then}~~~dk_{\cal K}~=~0~~,~~dk_{\cal H}~=~k_{\cal K}\wedge k_{\cal K} ~~~\label{noethermc}~~~.
\eea 
These are the equations \bref{MCflat},
and still represent the Poincar$\acute{\rm e}$ algebra 
but in the different gauge.
We will use 
the algebra obtained in \bref{noethermc}
 to construct nonlocal conserved charges.
Our criterion is conservation of charges so far.
At the end the obtained charges 
give the correct algebra after the quantization
in any case.

\par

\vskip 6mm

\subsection{ Nonlocal charges}

If we introduce the algebra $\left[T_A,T_B\right]=F_{AB}^CT_C$
corresponding to the 
 structure coefficient induced by the Noether current
 $F_{AB}^C$ in \bref{MCflat} and \bref{dJJJ},
then the covariant derivative operator can be defined as
\bea
{\cal D}_\alpha=\partial_\alpha-J^A_\alpha T_A~~\to~~
\left[\partial^\alpha, {\cal D}_\alpha\right]=0
~~{\rm and}~~
\left[{\cal D}_\alpha, {\cal D}_\beta\right]=0~~~.\label{DDD}
\eea
These equations correspond to 
the current conservation law \bref{cons} and the ``flatness" condition \bref{MCflat}
respectively.
Then we follow the BIZZ procedure 
using with this covariant derivative operator in \bref{DDD}.
An infinite set of conserved nonlocal currents are obtained
by the covariant derivative ${\cal D}_\alpha$  acting on the dual 
potential $\chi_{[N]}$'s 
\bea
{\cal J}_{[N]\alpha}^{~~A}=\epsilon_{\alpha\beta}\partial^\beta\chi_{[N]}^{~A}~~,~~
{\cal J}_{[N+1]\alpha}^{~~~~A}=({\cal D}_\alpha \chi_{[N]})^A~~\to
~~\partial^\alpha{\cal J}_{[N+1]\alpha}^{~~~~A}=0,~~_{N \geq 0}~~.\label{Brezin}
\eea  
Since $\chi_{[N]}$'s are nonlocal, obtained conserved currents
 ${\cal J}_{[N]}$ are also nonlocal.

Now we will compute the nonlocal charges for a string in the flat background.
We list expressions of zero-th, first and second order nonlocal charges here:
\bea
{\cal J}_{[0]}{}_\alpha^A&=&J^A_\alpha=\left\{
\begin{array}{l}
\T\partial_\alpha X^\mu\\
\T X^{[\mu}\partial_\alpha X^{\nu]} 
\end{array}\right.~~~\\
{\cal J}_{[1]}{}_\alpha^A&=&\left\{
\begin{array}{l}
\T\epsilon_{\alpha\beta} \partial^\beta X^\mu\\
\T\epsilon_{\alpha\beta}  X^{[\mu}\partial^\beta X^{\nu]}
-\partial_\alpha X^{[\mu} \chi_{[0]}^{\nu]} 
\end{array}\right.~~~\\
{\cal J}_{[2]}{}_\alpha^A&=&\left\{
\begin{array}{l}
 \T\partial_\alpha X^\mu\\
\T X^{[\mu}\partial_\alpha X^{\nu]}
-(\T)^{-1}\left(
J^{[\mu}_{[1]\alpha}\chi^{\nu]}_{[0]}
+J^{[\mu}_{[0]\alpha}\chi^{\nu]}_{[1]}
\right)
\end{array}\right.~~~
\eea
where $\epsilon^{\alpha\beta}\epsilon_{\beta\gamma}=\delta^\alpha_\gamma$
and $\epsilon^{\tau\sigma}=1=\epsilon_{\sigma\tau}$.

Let us consider a  closed string winding around some compact directions as
\bea
&X^\mu(\sigma+2\pi)=X^\mu(\sigma)+2\pi R w^\mu~~,~~p^\mu=\displaystyle\frac{n^\mu}{R}~~,~~
w^\mu, n^\mu\in{\bf Z}~~,\label{winding}
\eea
and it is quantized as
\bea
&X^\mu=x^\mu
+\alpha'p^\mu\tau+w^\mu R\sigma 
+\sqrt{\frac{\alpha'}{2}}\displaystyle\sum_{n\neq 0}
\frac{i}{n}\left(
\alpha_n^\mu e^{-in(\tau+\sigma)}+
\tilde{\alpha}_n^\mu e^{-in(\tau-\sigma)}
\right)~&\nn~~~.
\eea
It is required to write $x^\mu=x^\mu_+ +x^\mu_-$
in such a way that they become canonical conjugates of
momenta and windings,  
$
\left[x^\mu_+,\left(p^\nu+{w^\nu R}/{\alpha'} \right)/2\right]=i\eta^{\mu\nu}
=\left[{x}^\mu_-,\left(p^\nu-{w^\nu R}/{\alpha'}\right)/2 \right]$.
It is also noted that there is no restriction on
$x^\mu_+ -x^\mu_-$ at this stage.

For the finite range of $\sigma$ coordinate $0\le \sigma\le 2\pi$
with the nontrivial winding \bref{winding},
a conserved  charge is obtained by the integration along the 
following  path (Figure 1):
\begin{figure}[t]
\begin{center}
~~~~~~~\includegraphics[width=8cm,clip]{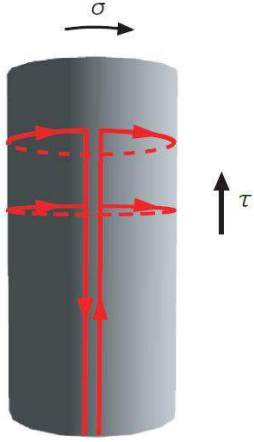}
\end{center}
\caption{The integration path.}
\end{figure}
\bea
Q_{[N]}^A=\displaystyle\int_{-\infty}^{\tau} d\tau' {\cal J}_{[N]} {}_\sigma^A(\tau',\sigma=0)
+\displaystyle\int_0^{2\pi} d\sigma {\cal J}_{[N]} {}_\tau^A(\tau,\sigma)
+\displaystyle\int_\tau^{-\infty} d\tau' {\cal J}_{[N]} {}_\sigma^A(\tau',\sigma=2\pi)\nn\\
\label{intpath}
\eea
and the dual potential $\chi$ defined in \bref{Brezin} 
is computed as
\bea
\chi^A(\tau,\sigma)=-\displaystyle\int_0^{\sigma} d\sigma'~J^A_\tau (\tau,\sigma')-
\displaystyle\int_{-\infty}^{\tau} d\tau'~J^A_\sigma(\tau',\sigma=0)~~~.
\eea

The resultant  Noether charges, which are zero-the order charges, 
 \bref{flatPM0} are again listed here:
\bea
Q_{[0]}^\mu=\hat{P}^\mu =p^\mu~~,~~
Q_{[0]}^{\mu\nu}=\hat{M}^{\mu\nu}=x^{[\mu}p^{\nu]}
+\frac{i}{2}\displaystyle\sum_{n\neq 0}\frac{1}{n}\left(
\alpha_n^{[\mu}\alpha_{-n}^{\nu]}
+\tilde{\alpha}_n^{[\mu}\tilde{\alpha}_{-n}^{\nu]}
\right)\label{Q0Q0}
\eea
The first order nonlocal charges:
\bea
&Q_{[1]}^\mu=-\omega^\mu R~~,~~
Q_{[1]}^{\mu\nu}=-x^{[\mu} \omega^{\nu]} R
-{i}\displaystyle\sum_{n\neq 0}\frac{1}{n}\left(
\alpha_n^{[\mu}\alpha_{-n}^{\nu]}
-\tilde{\alpha}_n^{[\mu}\tilde{\alpha}_{-n}^{\nu]}
\right)\label{Q1Q1}&
\eea
The even order,  $2N$-th ($N\ge 1$) order nonlocal charges:
\bea
Q_{[2N]}^\mu&=&Q_{[0]}^\mu \label{fleve}
\\
Q_{[2N]}^{\mu\nu}&=&Q_{[0]}^{\mu\nu}+N \Delta Q^{\mu\nu}_{[{\rm even}]}~~,~~
\Delta Q^{\mu\nu}_{[{\rm even}]}=i\displaystyle\sum_{n\neq 0}\frac{1}{n}\left(
\alpha_n^{[\mu}\alpha_{-n}^{\nu]}
+\tilde{\alpha}_n^{[\mu}\tilde{\alpha}_{-n}^{\nu]}
\right)\nn
\eea
and for the odd order, $2N+1$-th ($N\ge 1$) order nonlocal charges:
\bea
Q_{[2N+1]}^\mu&=&Q_{[1]}^\mu~~\label{flodd}\\
Q_{[2N+1]}^{\mu\nu}&=&Q_{[1]}^{\mu\nu}+N\Delta Q^{\mu\nu}_{[{\rm odd}]}~~,~~
\Delta Q^{\mu\nu}_{[{\rm odd}]}=-{i}\displaystyle\sum_{n\neq 0}\frac{1}{n}\left(
\alpha_n^{[\mu}\alpha_{-n}^{\nu]}
-\tilde{\alpha}_n^{[\mu}\tilde{\alpha}_{-n}^{\nu]}
\right)\nn
\eea
Independent components are 
zero-th and first order nonlocal charges for zero mode 
and non-zero mode separately.
The momentum, the winding number, the 
total Lorentz spin  and the relative Lorentz spin, 
which is the difference between the left mover spin and  the right mover spin,
for zero mode
and for non-zero modes are conserved separately.

\par
\vskip 6mm

\subsection{ T-duality}

T-duality transformation interchanges the momentum and the winding 
 by interchanging $R\leftrightarrow 1/R$.
T-duality transformation in $X(\tau,\sigma)$ reduces 
the interchange $\tau\leftrightarrow \sigma$ 
which causes 
\bea
n^\mu\leftrightarrow w^\mu~~,~~R\leftrightarrow 1/R~~,~~
\tilde{\alpha}_n^\mu\leftrightarrow -\tilde{\alpha}_{-n}^\mu~~~.
\eea

The Noether charges after the T-duality transformation is
given
\bea
Q_{[0]}^\mu=\tilde{P}^\mu=w^\mu R~~,~~
Q_{[0]}^{\mu\nu}=\tilde{M}^{\mu\nu}=x^{[\mu} w^{\nu]} R
+\frac{i}{2}\displaystyle\sum_{n\neq 0}\frac{1}{n}\left(
\alpha_n^{[\mu}\alpha_{-n}^{\nu]}
-\tilde{\alpha}_n^{[\mu}\tilde{\alpha}_{-n}^{\nu]}
\right)~~~.
\eea  
These Noether charges 
correspond to the first order nonlocal charges in the original background
\bref{Q1Q1}.
Further higher order nonlocal charges after T-duality transformation are:
\bea
&&Q_{[1]}^\mu=-p^\mu~~,~~
Q_{[1]}^{\mu\nu}=-x^{[\mu} p^{\nu]}
-{i}\displaystyle\sum_{n\neq 0}\frac{1}{n}\left(
\alpha_n^{[\mu}\alpha_{-n}^{\nu]}
+\tilde{\alpha}_n^{[\mu}\tilde{\alpha}_{-n}^{\nu]}
\right)\label{Q1Q1T}\\
&&Q_{[2N]}^\mu=Q_{[0]}^\mu~~,~~
Q_{[2N]}^{\mu\nu}=Q_{[0]}^{\mu\nu}+N\Delta Q^{\mu\nu}_{[{\rm even}]}~~,~~
\Delta Q^{\mu\nu}_{[{\rm even}]}=i\displaystyle\sum_{n\neq 0}\frac{1}{n}\left(
\alpha_n^{[\mu}\alpha_{-n}^{\nu]}
-\tilde{\alpha}_n^{[\mu}\tilde{\alpha}_{-n}^{\nu]}
\right)~~\nn\\
&&Q_{[2N+1]}^\mu=Q_{[1]}^\mu~~,~~
Q_{[2N+1]}^{\mu\nu}=Q_{[1]}^{\mu\nu}+N \Delta Q^{\mu\nu}_{[{\rm odd}]}~~,~~
\Delta Q^{\mu\nu}_{[{\rm odd}]}=-{i}\displaystyle\sum_{n\neq 0}\frac{1}{n}\left(
\alpha_n^{[\mu}\alpha_{-n}^{\nu]}
+\tilde{\alpha}_n^{[\mu}\tilde{\alpha}_{-n}^{\nu]}
\right)\nn
\eea
The first order nonlocal charges in the T-dualized background correspond to
the zero-th order charges of the original background \bref{Q0Q0}.
The even and the odd order nonlocal charges are interchanged by T-duality.
The set of independent conserved charges is equivalent before and 
after the T-duality transformation in a flat space background.

\par
\vskip 6mm
\section{ Type IIB pp-wave background}

In this section we will 
compute the nonlocal charges for a closed string 
in the pp-wave background explicitly.
The isometry algebra is inhomogeneous SO($n$),
so we apply our generalized BIZZ procedure to obtain
nonlocal charges.
Before introducing nontrivial winding modes for the 
T-duality transformation
we present  the nonlocal charges 
in the type IIB pp-wave background.

\par
\vskip 6mm

\subsection{ Noether charges}

The type IIB pp-wave background is given by 
\bea
ds^2&=&2dX^+dX^-+dX^idX^i-4\mu^2X^iX^i(dX^+)^2~~,~~_{i=1,\cdots,8}\nn\\
^{(5)}F&=&\mu dX^+\left(
dX^1dX^2dX^3dX^4+dX^5dX^6dX^7dX^8
\right)\label{metricIIB}~~~.
\eea
The action for a string in this background is
\bea
&S=-
\displaystyle\frac{1}{4\pi\alpha'}\displaystyle\int~d^2\sigma
\left[2\partial_\alpha X^+\partial^\alpha X^-
-4{\mu}^2 {X}^iX^i\partial_\alpha X^+\partial^\alpha X^+
+\partial_\alpha X^i\partial^\alpha X^i
\right]~~&
~\nn\\&~~~~~~~~~~~~~~~~~
~~~~~~~~~~~~~~~~~~~~~~~~~~~~_{i=1,\cdots,8}~~~.&
\eea
Noether currents of the system are given by:
\bea
\left\{\begin{array}{ccl}
J^+_\alpha&=& \frac{1}{2\pi}{\partial_\alpha}X^+\\
J^{-}_\alpha&=& \frac{1}{2\pi\alpha'}\left({\partial_\alpha}X^--4\mu^2 X^iX^i {\partial_\alpha}X^+\right)\\
J^i_\alpha&=&\T\left( {\partial_\alpha}X^i\cos 2\mu X^+ -X^i{\partial_\alpha}\cos 2\mu X^+\right)\\
J^{i*}_\alpha&=&\frac{1}{4\pi\mu}\left( {\partial_\alpha}X^i\sin 2\mu X^+ -X^i{\partial_\alpha}
\sin 2\mu X^+\right)\\
J^{ij}_\alpha&=&\T X^{[i}{\partial_\alpha}X^{J]}
\end{array}\right.\label{NoetherppB}
\eea

In the lightcone gauge $X^+=p^+\tau$ 
the action becomes  
\bea
S=-\frac{1}{4\pi\alpha'}\displaystyle\int~d^2\sigma
\left[
\partial_\alpha X^i\partial^\alpha X^i
+\hat{\mu}^2 {X}^iX^i
\right]~~,~
~\hat{\mu}=2\mu p^+~~
,~~~_{i=1,\cdots,8}
\eea
and Virasoro constraints are
\bea
&&\left\{\begin{array}{l}
h=2{\cal P}^-p^++\dot{X}^i\dot{X}^i
+X^{i'}X^{i'}
+\hat{\mu}^2X^iX^i=0\\
t=p^+X^{-'}+\dot{X}^iX^{i'}=0
\end{array}\right.\label{Viralc}
\\
&&~{\cal P}^{-}=\dot{X}^--4\mu^2p^+X^iX^i ~~\nn~~~
\eea 
with $\dot{X}=\partial_\tau X$ and $X'=\partial_\sigma X$.
The quantization in the lightcone gauge is performed as
\bea
{ X}^i&=&{x}^i\cos \hat{\mu} \tau +{p}^i
\frac{\alpha'}{\hat{\mu}}
\sin \hat{\mu} \tau+i\sqrt{\displaystyle\frac{\alpha'}{2}}
\displaystyle\sum_{n\neq 0}
\frac{1}{n\rho_n}\left(
{\alpha}^i_ne^{-in\sigma}+\tilde{\alpha}^i_ne^{in\sigma}
\right)e^{-in\rho_n\tau}\nn\\
&&\rho_n=\sqrt{1+\left(\frac{\hat{\mu}}{n}\right)^2}
\eea
with
\bea
&&\left[x^i,p^j\right]=i\delta^{ij}~~,~~
\left[\alpha^i_n,\alpha^j_m\right]=n\rho_n\delta^{ij}\delta_{n,-m}
=\left[\tilde{\alpha}^i_n,\tilde{\alpha}^j_m\right]~~~.
\eea

Noether charges are given as
\bea
\left\{\begin{array}{ccccl}
Q_{[0]}^+&=&Q^+&=& p^+\\
Q_{[0]}^-&=&Q^{-}&=& -\frac{1}{2p^+}\left[
\alpha' p^ip^i+\displaystyle\frac{\hat{\mu}^2}{\alpha'} x^ix^i
+\displaystyle\sum_{n \neq 0} (\alpha_n^i \alpha_{-n}^i
+\tilde{\alpha}_n^i \tilde{\alpha}_{-n}^i)
\right]\\
Q_{[0]}^i&=&Q^i&=&p^i\\
Q_{[0]}^{i*}&=&Q^{i*}&=&-p^+x^i\\
Q_{[0]}^{ij}&=&Q^{ij}&=&x^{[i}p^{j]}+\frac{i}{2}
\displaystyle\sum_{n \neq 0}\frac{1}{n\rho_n} (\alpha_n^{[i} \alpha_{-n}^{j]}
+\tilde{\alpha}_n^{[i} \tilde{\alpha}_{-n}^{j]})
\end{array}\right.\label{NoetherQ0}
\eea
which satisfy the following pp-wave algebra:
\bea
\begin{array}{lclcl}
\left[Q^-,Q^i \right]=i\frac{4\mu^2}{\alpha'} Q^{i*}&,&
\left[Q^-,Q^{i*} \right]=-i \alpha' Q^{i}&,&
\left[Q^i,Q^{j*} \right]=i\delta^{ij} Q^{+}\\
\left[Q^{ij},Q^{k} \right]=-i\delta^{k[i} Q^{j]}&,&
\left[Q^{ij},Q^{k*} \right]=-i\delta^{k[i} Q^{j]*}&,&
\left[Q^{ij},Q^{kl} \right]=-i\delta^{[i|[l} Q^{k]|j]}\\
{\rm others}=0&&&&
\end{array}\label{ppIIB}
\eea

\par\vskip 6mm

\subsection{ Nonlocal charges}

One form currents constructed from 
the Noether currents \bref{NoetherppB} as $J=d\sigma^\alpha J_\alpha$
  satisfy the following  equations:
\bea
\left\{
\begin{array}{lcl}
dJ^+=0&,&
dJ^-=16\pi\mu^2 J^i{\wedge} J^{i*}\\
dJ^i=-\displaystyle\frac{16\pi\mu^2}{\alpha'} J^+{\wedge} J^{i*}&,&dJ^{i*}={4\pi\alpha'} J^+{\wedge} J^{i}
\\
dJ^{ij}=4\pi\alpha' \left(J^{i}{\wedge} J^{j}
+\displaystyle\frac{4\mu^2}{\alpha'^2} J^{i*}{\wedge} J^{j*}
\right)
\end{array}\label{IIBBNtmc}\right.
\eea
They are ``flatness" condition and the 
consistency condition $ddJ=0$ is also satisfied,
so the covariant derivative operator \bref{DDD} can be constructed.
The  structure constant read off from the above equations \bref{IIBBNtmc}
is different from the one of  the pp-wave algebra \bref{ppIIB}
as expected in the subsection 2.1. 

According to the procedure \bref{Brezin}
the first order nonlocal currents are given by
\bea
\left\{
\begin{array}{ccl}
{\cal J}_{[1]}{}^+_\alpha&=&\epsilon_{\alpha\beta}J^+{}^\beta\\
{\cal J}_{[1]}{}^-_\alpha&=&\epsilon_{\alpha\beta}J^-{}^\beta
-8\pi\mu^2\left(J^i_\alpha \chi_{[0]}^{i*}
-J^{i*}_\alpha \chi_{[0]}^{i}
\right)\\
{\cal J}_{[1]}{}^i_\alpha&=&\epsilon_{\alpha\beta}J^i{}^\beta
+\frac{8\pi\mu^2}{\alpha'}\left(J^+_\alpha \chi_{[0]}^{i*}
-J^{i*}_\alpha \chi_{[0]}^{+}
\right)\\
{\cal J}_{[1]}{}^{i*}_\alpha&=&\epsilon_{\alpha\beta}J^{i*}{}^\beta
-2\pi{\alpha'}\left(J^+_\alpha \chi_{[0]}^{i}
-J^{i}_\alpha \chi_{[0]}^{+}
\right)\\
{\cal J}_{[1]}{}^{ij}_\alpha&=&
\epsilon_{\alpha\beta}J^{ij}{}^\beta
-2\pi{\alpha'}
\left( J^{[i}_\alpha \chi_{[0]}^{j]}
+\displaystyle\frac{4\mu^2}{\alpha'^2} 
J^{[i*}_\alpha \chi_{[0]}^{j]*}
\right)
\end{array}\right.
\eea 
where the dual potential $\chi_{[0]}$'s are given as:
\bea
\chi_{[0]}^+&=&-\frac{1}{2\pi}p^+\sigma\\
\chi_{[0]}^i&=&-\frac{1}{2\pi}p^i\sigma-\frac{i}{4\pi\sqrt{2\alpha'}}
\displaystyle\sum_{n\neq 0}\frac{1}{n}
\left((1+\frac{\hat{\mu}}{n\rho_n})e^{i\hat{\mu}\tau}
+(1-\frac{\hat{\mu}}{n\rho_n})e^{-i\hat{\mu}\tau}
\right)
\left(\alpha_n^ie^{-in\sigma}
-\tilde{\alpha}_n^ie^{in\sigma
}
\right)
e^{-in\rho_n\tau}
\nn\\
\chi_{[0]}^{i*}&=&\frac{1}{2\pi}p^+x^i\sigma
-\frac{\sqrt{\alpha'}}{8\pi\mu\sqrt{2}}
\displaystyle\sum_{n\neq 0}\frac{1}{n}
\left((1+\frac{\hat{\mu}}{n\rho_n})e^{i\hat{\mu}\tau}
-(1-\frac{\hat{\mu}}{n\rho_n})e^{-i\hat{\mu}\tau}
\right)
\left(\alpha_n^ie^{-in\sigma}
-\tilde{\alpha}_n^ie^{in\sigma}
\right)
e^{-in\rho_n\tau}
\nn
\eea 
The first order nonlocal charges are obtained as:
\bea
\left\{\begin{array}{ccl}
Q_{[1]}^-&=&\displaystyle\frac{1}{2p^+}\displaystyle\sum_{n\neq 0}
\frac{1}{\rho_n}\left(1+2\left(\frac{\hat{\mu}}{n}\right)^2\right)
\left(\alpha_n^i\alpha_{-n}^i-\tilde{\alpha}_n^i\tilde{\alpha}_{-n}^i\right)\\
Q_{[1]}^{ij}&=&i\displaystyle\sum_{n\neq 0}
\frac{1}{n}
\left(\alpha_n^{[i}\alpha_{-n}^{j]}
-\tilde{\alpha}_n^{[i}\tilde{\alpha}_{-n}^{j]}
\right)\\
{\rm others}&=&0
\end{array}\right.
\eea
where $X^{-'}$ in ${\cal J}_{[1]\tau}^{~-}$ 
is determined from the Virasoro condition $t=0$ in \bref{Viralc}.
In the ref. \cite{Alday:2003zb} this term was set to be zero because of
the total derivative of $\sigma$, but
there is a non-zero contribution 
determined by the Virasoro condition.

The second order nonlocal current is given by
\bea
{\cal J}_{[2]}{}^A_\alpha&=&
J^A_\alpha-\frac{1}{2}F^A_{BC}\left(
\epsilon_{\alpha\beta}J^{B\beta}\chi^C_{[0]}+J^{B}_\alpha \chi_{[1]}^C
\right)
\eea
and the second order nonlocal charges are obtained as
\bea
\left\{\begin{array}{ccl}
Q_{[2]}^--Q_{[0]}^-&=&-\displaystyle\frac{1}{p^+}\displaystyle\sum_{n\neq 0}
\left(\frac{\hat{\mu}}{n\rho_n}\right)^2\left(1+2\left(\frac{\hat{\mu}}{n}\right)^2\right)
\left(\alpha_n^i\alpha_{-n}^i+\tilde{\alpha}_n^i\tilde{\alpha}_{-n}^i\right)\\
Q_{[2]}^{ij}-Q_{[0]}^{ij}&=&i\displaystyle\sum_{n\neq 0}
\frac{1}{n\rho_n}\left(1+2\left(\frac{\hat{\mu}}{n}\right)^2
\right)
\left(\alpha_n^{[i}\alpha_{-n}^{j]}
+\tilde{\alpha}_n^{[i}\tilde{\alpha}_{-n}^{j]}
\right)\\
{\rm others}&=&0
\end{array}\right.\label{Q2Q2}
\eea
These charges \bref{Q2Q2} are different from $Q_{[0]}$ in \bref{NoetherQ0}.
Higher order nonlocal charges are obtained by this
computation by the procedure \bref{Brezin},
and it is expected that
coefficients on the nonlocal charges
are different in each order unlike the flat case \bref{fleve} and \bref{flodd}.
There exist 
an infinite number of independent  conserved charges exist
 in the pp-wave background contrast to the flat case.

\par
\vskip 6mm
\section{ T-dual pp-wave backgrounds}

Now we will examine the nonlocal charges for closed strings
with a winding mode in the pp-wave backgrounds.
We will compare the nonlocal charges 
before  and after the T-duality transformation.

\par
\subsection{ Michelson's cycle}

In order to examine T-duality a spacelike circle 
is needed to compactify it.
First we rewrite IIB pp-wave coordinate in 
\bref{metricIIB} in terms of $x^\mu$,
then change variables as
\bea
\left\{\begin{array}{ccl}
x^+&=&X^+\\
x^-&=&X^--2\mu X^1X^2\\
x^I&=&X^I~~,~~_{I=3,\cdots,8}\\
\left(
\begin{array}{c}x^1\\x^2\end{array}
\right)&=&
\left(
\begin{array}{cc}\cos 2\mu X^+&-\sin 2\mu X^+ \\\sin 2\mu X^+& \cos 2\mu X^+\end{array}
\right)\left(
\begin{array}{c}X^1\\X^2\end{array}
\right)
\end{array}\right.
\eea
to get
\bea
ds^2&=&2dX^+dX^--4\mu^2X^IX^I(dX^+)^2+dX^idX^i-8\mu  
dX^+X^2dX^1~~,~~_{i=1,\cdots,8~~~I=3,\cdots,8}\nn\\
^{(5)}F&=&\mu dX^+\left(
dX^1dX^2dX^3dX^4+dX^5dX^6dX^7dX^8
\right)~~~.\label{Michel12}
\eea
The action of a string in the Michelson's pp-wave geometry 
 is given by
\bea
&S_{IIB}=-\displaystyle\frac{1}{4\pi\alpha'}\displaystyle\int d^2\sigma 
\left[2\partial_\alpha X^+ \partial^\alpha X^-
-4{\mu}^2X^I X^I\partial_\alpha X^+ \partial^\alpha X^+
+\partial_\alpha X^i \partial^\alpha X^i
-8{\mu} X^2 \partial_\alpha {X}^1\partial^\alpha X^+
\right]&\nn\\
&~~~_{i=1,2,3,\cdots,8}~~,~~_{I=3,\cdots,8}
~~~.&
\eea
Target space indices $-,1,2$ are renamed in such 
a way that they coincide 
translations and rotations of
the new coordinate basis \bref{Michel12}.
We compactify a spacelike S$^1$ along the cycle $S_{12}^+$ 
whose isometry $k_{S_{12}^+}=
k_{e_1}+\frac{1}{2\mu}k_{e_2^*}$.
It breaks symmetry from 
SO(4)$\times$SO(4)  into  SO(2)$\times$SO(4), so $1+6$  isometries survive.
Some translations and boosts are also broken,
and we list  survived isometries 
in Michelson's notation
 corresponding 
to our notation up to  normalization:
\bea
&\begin{array}{rcl}
k_{e_-}&{\Leftrightarrow}&\fr{k}
^+=\frac{\partial}{\partial X^-}\\
k_{e_+}-2\mu k_{M_{12}}&{\Leftrightarrow}&\fr{k}^-=\frac{\partial}{\partial X^+}\\
k_{e_1}+\frac{1}{2\mu}k_{e_2^*}=k_{S^{+}_{12}}&{\Leftrightarrow}&\fr{k}^1=\frac{\partial}{\partial X^1}\\
k_{e_2}+\frac{1}{2\mu}k_{e_1^*}=k_{S^{+}_{21}}&{\Leftrightarrow}&\fr{k}^{2}=
(\cos 4\mu X^+)\frac{\partial}{\partial X^2}
+(\sin 4\mu X^+)\frac{\partial}{\partial X^1}
+4\mu(\sin 4\mu X^+)X^2 \frac{\partial}{\partial X^-}
\\
k_{e_1}-\frac{1}{2\mu}k_{e_2^*}=k_{S^{-}_{12}}&{\Leftrightarrow}&\fr{k}^{2*}
=(\cos 4\mu X^+)\frac{\partial}{\partial X^1}
-(\sin 4\mu X^+)\frac{\partial}{\partial X^2}
+4\mu(\cos 4\mu X^+)X^2 \frac{\partial}{\partial X^-}
\\
k_{e_I}&{\Leftrightarrow}&\fr{k}^I=
(\cos 2\mu X^+)\frac{\partial}{\partial X^I}
+2\mu(\sin 4\mu X^+)X^I \frac{\partial}{\partial X^-}
\\
k_{e_I^*}&{\Leftrightarrow}&\fr{k}^{I*}=
(\sin 2\mu X^+)\frac{\partial}{\partial X^I}
-2\mu(\cos 4\mu X^+)X^I \frac{\partial}{\partial X^-}
~~\\
&&\end{array}\nn&\\&&
\eea
A Killing vector, 
\bea
&k_{S^-_{21}}=k_{e_2}-\frac{1}{2\mu}k_{e_1^*}\Leftrightarrow \fr{k}^{1*}
=\frac{\partial}{\partial X^2}
+4\mu X^1\frac{\partial}{\partial X^-}~~,~~~~~~~~~~~~~~~~~~~~~~~~~~~~~~~~~~~~~~~~~~~~&
\eea
 becomes
multivalued after compactification of $X^1$ direction
so ill-defined.
There are $17$ isometries survived and 
they are noncompact Killing vectors except $k_{S^{+}_{12}}$. 
This geometry is totally  
$1+6+17=24$ dimensional group. 
The Noether currents, $J^A_\alpha$, corresponding to the Killing vectors $\fr{k}^A$ 
are given as:
\bea
\left\{\begin{array}{ccl}
J^+_\alpha&=&\frac{1}{2\pi}{\partial_\alpha}X^+\\
J^-_\alpha&=&\frac{1}{2\pi\alpha'}\left({\partial_\alpha}X^-  -4\mu^2X^IX^I{\partial_\alpha}X^+ -4\mu X^2{\partial_\alpha}X^1 \right)\\
J^1_\alpha&=&\frac{1}{2\pi\alpha'}\left({\partial_\alpha}X^1-4\mu X^2{\partial_\alpha}X^+\right)\\
J^{1*}_\alpha&=&\frac{1}{4\pi\mu}\left({\partial_\alpha}X^2+4\mu X^1{\partial_\alpha}X^+\right)\\
J^{2}_\alpha&=&\frac{1}{2\pi\alpha'}\left({\partial_\alpha}X^2~
\cos 4\mu X^++ {\partial_\alpha}X^1 ~\sin 4\mu X^+ \right)\\
J^{2*}_\alpha&=&\frac{1}{4\pi\mu}\left({\partial_\alpha}X^2~ \sin 4\mu X^+
-{\partial_\alpha}X^1 ~\cos 4\mu X^+\right)\\
J^I_\alpha&=&\frac{1}{2\pi\alpha'}\left({\partial_\alpha}X^I ~\cos 2\mu X^+ -X^I{\partial_\alpha}\cos 2\mu X^+\right)\\
J_\alpha^I{}^*&=&\frac{1}{4\pi\mu}\left({\partial_\alpha}X^I~ \sin 2\mu X^+ -X^I{\partial_\alpha}\sin 2\mu X^+\right)\\
J^{IJ}_\alpha&=&\frac{1}{2\pi\alpha'}X_{[I}{\partial_\alpha}X_{J]}
\end{array}\right.\label{S12JJ}
\eea
where $J^{1*}_\alpha$ is the one for the non-compact case
although it will be compactified soon later.

In the lightcone gauge
the action for a string in the Michelson's pp-wave geometry  becomes 
\bea
S_{IIB}&=&-\frac{1}{4\pi\alpha'}\int d^2\sigma ~
\left[
\partial_\alpha X^i \partial^\alpha X^i
+\hat{\mu}^2X^I X^I
+4\hat{\mu} X^2 \dot{X}^1
\right]\nn\\
&&~~~\quad\quad \hat{\mu}=2p^+\mu ~~,~~_{i=1,2,3,\cdots,8}~~,~~_{I=3,\cdots,8}
~~~,
\eea
and the Virasoro constraints are
\bea
&&\left\{\begin{array}{ccl}
h_{IIB}&=&2{\cal P}^- p^++\dot{X}^i\dot{X}^i+{X}^{i'}{X}^{i'}
+\hat{\mu}^2X^IX^I=0\\
t_{IIB}&=&p^+ X^{-'}+\left(\dot{X}^1-2\hat{\mu} X^2\right)X^{1'}
+\dot{X}^2 X^{2'}+\dot{X}^I X^{I'}=0\\
\end{array}\right.\\
&&~~~{\cal P}^- =\dot{X}^- -4\mu^2 p^+X^IX^I-4\mu X^2\dot{X}^1\nn~~~.
\eea

A closed string  with a winding mode $X^1(\sigma+2\pi)=X^1(\sigma)+2\pi Rw$ 
can be quantized as  
\bea
&\begin{array}{rcl}
X^I&=&x^I\cos \hat{\mu}\tau+\frac{\alpha'}{\hat{\mu}}p^I\sin \hat{\mu}\tau
+i\sqrt{\frac{\alpha'}{2}}\displaystyle\sum_{n \neq 0}
\frac{1}{n\rho_n}\left(
\alpha^I_ne^{-in\sigma}+
\tilde{\alpha}^I_ne^{+in\sigma}
\right)
e^{-in\rho_n\tau}
\nn\\
X^1+iX^2&=&wR\sigma +e^{-i\hat{\mu}\tau}
\left\{x\cos \hat{\mu}\tau+\frac{\alpha'}{\hat{\mu}}p\sin \hat{\mu}\tau
+i\sqrt{\frac{\alpha'}{2}}\displaystyle\sum_{n \neq 0}
\frac{1}{n\rho_n}\left(
\alpha_ne^{-in\sigma}+
\tilde{\alpha}_ne^{in\sigma}
\right)e^{-in\rho_n\tau}
\right\}
\nn\\
X^1-iX^2&=&w R\sigma +e^{i\hat{\mu}\tau}
\left\{\bar{x}\cos \hat{\mu}\tau+\frac{\alpha'}{\hat{\mu}}\bar{p}\sin \hat{\mu}\tau
+i\sqrt{\frac{\alpha'}{2}}\displaystyle\sum_{n \neq 0}
\frac{1}{n\rho_n}\left(
\bar{\alpha}_ne^{-in\sigma}+
\tilde{\bar{\alpha}}_ne^{in\sigma}
\right)e^{-in\rho_n\tau}
\right\}
\end{array}
&\nn\\
&\rho_n=\sqrt{1+\frac{\hat{\mu}^2}{n^2}}~~~~~~~~~~~~~~~~~~~~~~~~~&\label{IIBXXX}
\eea
with
\bea
&\left[x^I,p^J\right]=i\delta^{IJ}~~,~~
\left[x,\bar{p}\right]=i=\left[\bar{x},p\right]&\nn\\
&\left[\alpha_n^I,\alpha_m^J\right]=n\rho_n\delta^{IJ}\delta_{n,-m}
=\left[\tilde{\alpha}_n^I,\tilde{\alpha}_m^J\right]&\\
&\left[\alpha_n,\bar{\alpha}_m\right]=2n\rho_n\delta_{n,-m}
=\left[\tilde{\alpha}_n,\tilde{\bar{\alpha}}_m\right]\nn~~~.
\eea

The Noether charges,  which are the zero-th order charges,
 are followings:
\bea
\left\{\begin{array}{ccl}
Q_{[0]}^+&=&p^+\\
Q_{[0]}^-&=&
-\frac{1}{2p^+}\left[
\displaystyle\sum_{i=2,3,\cdots,8}\left(\alpha'p^ip^i+\frac{\hat{\mu}^2}{\alpha'}x^ix^i\right)
+\frac{1}{\alpha'}(wR)^2\right.\\&&\left.
+\displaystyle\sum_{n\neq 0}
\left\{
(\alpha_n^I\alpha_{-n}^I+\tilde{\alpha}_n^I\tilde{\alpha}_{-n}^I)
+\left(1+\frac{\hat{\mu}}{n\rho_n}\right)
\left(
\alpha_n\bar{\alpha}_{-n}+\tilde{{\alpha}}_n\tilde{\bar{\alpha}}_{-n}
\right)
\right\}
\right]\\
Q_{[0]}^1&=&p^1\\
Q_{[0]}^{1*}&=&
p^+\left(x^1+2\pi w R\right)\\
Q_{[0]}^2&=&p^2\\
Q_{[0]}^{2*}&=&p^+x^2\\
Q_{[0]}^I&=&p^I\\
Q_{[0]}^I{}^*&=&-p^+x^I\\
Q_{[0]}^{IJ}&=&-x^{[I}p^{J]}+\frac{i}{2}\displaystyle\sum_{n\neq 0}
\displaystyle\frac{1}{n\rho_n}
\left(\alpha_n^{[I}\alpha_{-n}^{J]}+\tilde{\alpha}_n^{[I}\tilde{\alpha}_{-n}^{J]}
\right)\end{array}\right.\label{IIBNoether}
\eea 
where zero mode variables are rewritten 
as
\bea
&&\left\{\begin{array}{lcl}
p^1=\frac{1}{2}(p+\bar{p})+i\frac{\hat{\mu}}{2\alpha'}(x-\bar{x})&,&
x^1=\frac{1}{2}(x+\bar{x})-i\frac{\alpha'}{2\hat{\mu}}(p-\bar{p})\\
p^2=\frac{1}{2i}(p-\bar{p})-\frac{\hat{\mu}}{2\alpha'}(x+\bar{x})&,&
x^2=\frac{i}{2}(x-\bar{x})-\frac{\alpha'}{2\hat{\mu}}(p+\bar{p})
\end{array}\right.\\
&&\left[x^1,p^1\right]=i=\left[x^2,p^2\right]~~~.
\eea
The above charges excluding $Q^1$ and $Q^{1*}$ make 
a closed algebra.
After compactify $X^1$ direction as
$X^1\sim X^1+2\pi wR$ 
the winding mode breaks $Q_{[0]}^{1*}$ symmetry of the vacuum.

The one form currents made from the Noether currents in \bref{IIBNoether} as
$J^A=d\sigma^\alpha J_\alpha^A$ satisfy
the following M.C. equations:
\bea
\left\{
\begin{array}{lcl}
dJ^+=0&,&
dJ^-=16\pi\mu^2 \left(J^2{\wedge} J^{2*}+J^I{\wedge} J^{I*}\right)\\
dJ^1=\frac{16\pi\mu^2}{\alpha'} J^+{\wedge} J^{1*}&,&
dJ^{1*}=-4\pi\alpha' J^{+}{\wedge} J^1\\
dJ^2=-\frac{16\pi\mu^2}{\alpha'}J^+{\wedge} J^{2*}&,&
dJ^{2*}=4\pi\alpha' J^{+}{\wedge} J^2\\
dJ^I=-\frac{16\pi\mu^2}{\alpha'} J^+{\wedge} J^{I*}&,&dJ^{I*}={4\pi\alpha'} J^+{\wedge} J^{I}
\\
dJ^{IJ}=4\pi\alpha' \left(J^{I}{\wedge} J^{J}
+\frac{4\mu^2}{\alpha'^2} J^{I*}{\wedge} J^{J*}
\right)
\end{array}\label{MCIIBS1}\right.
\eea
The structure constant 
read off from these M.C. equations is different from  the algebra generated by
charges \bref{IIBNoether}
again.
The consistency $ddJ=0$ corresponding to the Jacobi identity 
is satisfied without $J^1$ and $J^{1*}$.
Then compactification of $X^1$-direction is possible consistently.
The first order conserved nonlocal charges are obtained as follows:
\bea
\left\{\begin{array}{ccl}
Q_{[1]}^1&=&-\frac{wR}{\alpha'}\\
Q_{[1]}^-&=&\frac{1}{2p^+}\displaystyle\sum_{n\neq 0}
\frac{1}{\rho_n}\left\{\left(1+2\left(\frac{\hat{\mu}}{n}\right)^2\right)
(\alpha_n^I\alpha_{-n}^I-\tilde{\alpha}_n^I\tilde{\alpha}_{-n}^I)
\right.
\\&&~~~~~~~~~~~~~~~~~~~~
\left.
+\left(\rho_n+\displaystyle\frac{\hat{\mu}}{n}\right)^2
\left(
\alpha_n\bar{\alpha}_{-n}-\tilde{{\alpha}}_n\tilde{\bar{\alpha}}_{-n}
\right)
\right\}\\
Q_{[1]}^{ij}&=&i\displaystyle\sum_{n\neq 0}
\frac{1}{n}
\left(\alpha_n^{[i}\alpha_{-n}^{j]}
-\tilde{\alpha}_n^{[i}\tilde{\alpha}_{-n}^{j]}
\right)\\
{\rm others}&=&0
\end{array}\right.\label{IIBnnc}
\eea
It is essential to use the integration path in \bref{intpath}
to obtain consistent charges especially $Q^-_{[1]},Q^2_{[1]},Q^{2*}_{[1]}$.

\par\vskip 6mm

\subsection{ T-duality}
T-duality of the type IIB nine dimensional geometry \bref{Michel12}
along $X^1$-direction 
transforms  into the type IIA
background . 
For the pp-wave background  
the NS/NS two-form potential $B$ 
and the R/R three form potential$^{(3)}C$
of the type IIA theory
are 
responsible for the 
nontrivial geometry the type IIB theory,
\bea
ds^2&=&2dX^+dX^-
-4\mu^2\left(X^IX^I+4(X^2)^2\right)(dX^+)^2+dX^idX^i
~~,~~_{i=1,2,\cdots,8,~~~I=3,\cdots,8}\nn\\
^{(3)}C&=&8\mu X^+ dX^2dX^3dX^4 ~~,~~
B~=~-4\mu X^2 dX^1dX^+~~.
\eea
The action for a string in this IIA background is
\bea
S_{IIA}&=&-\frac{1}{4\pi\alpha'}\int d^2\sigma ~
\left[2\partial_\alpha X^+ \partial^\alpha X^-
-4{\mu}^2\left(X^I X^I+4(X^2)^2\right)\partial_\alpha X^+\partial^\alpha X^+\right.
\nn\\
&&\left.~~~~~~~~~~~~~~~~~~+\partial_\alpha X^i \partial^\alpha X^i
+4{\mu}X^2\epsilon^{\alpha\beta}\partial_\alpha X^1\partial_\beta X^+
\right]~~~.\label{actIIA}
\eea
The Noether currents in the IIA pp-wave background 
are given as:
\bea
\left\{\begin{array}{ccl}
J^+_\alpha&=&\frac{1}{2\pi}\partial_\alpha X^+\\
J^-_\alpha&=&\frac{1}{2\pi\alpha'}\left\{\partial_\alpha X^-  
-4\mu^2\left(X^IX^I+4(X^2)^2\right) \partial_\alpha X^+ 
-2\mu X^2 \epsilon_{\alpha\beta}\partial^\beta X^1 \right\}\\
J^1_\alpha&=&\frac{1}{2\pi\alpha'}\left(\partial_\alpha X^1+4\mu X^2
\epsilon_{\alpha\beta} \partial^\beta X^+\right)\\
J^{1*}_\alpha&=&\frac{1}{2\pi}\left(X^+\partial_\alpha X^1
-X^1\partial_\alpha X^+ +2\mu X^2\epsilon_{\alpha\beta}\partial^\beta
(X^+)^2\right)\\
J^{2}_\alpha&=&\frac{1}{4\pi\alpha'}\left(\partial_\alpha X^2~
\cos 4\mu X^+ -X^2 
\partial_\alpha \cos 4\mu X^+ \right)\\
J^{2*}_\alpha&=&-\frac{1}{8\pi\mu}\left(\partial_\alpha X^2~\sin 4\mu X^+-
X^2 \partial_\alpha \sin 4\mu X^+\right)\\
J^I_\alpha&=&\frac{1}{2\pi\alpha'}\left(\partial_\alpha X^I
~ \cos 2\mu X^+-X^I\partial_\alpha \cos 2\mu X^+ \right)\\
J^I{}^*_\alpha&=&\frac{1}{4\pi\mu}\left(\partial_\alpha X^I
~ \sin 2\mu X^+ -X^I \partial_\alpha \sin 2\mu X^+ \right)\\
J^{IJ}_\alpha&=&\frac{1}{2\pi\alpha'}X^{[I}\partial_\alpha X^{J]}
\end{array}\right.\label{S12JJ}
\eea

In the lightcone gauge the action becomes
\bea
S_{IIA}=-\frac{1}{4\pi\alpha'}\int d^2\sigma ~
\left[
\partial_\alpha X^i \partial^\alpha X^i
-\hat{\mu}^2\left(X^I X^I+4(X^2)^2\right)
-4\hat{\mu}X^2X^1{}'
\right]~~~,\label{actIIAlc}
\eea
and the Virasoro constraints are
\bea
&&\left\{\begin{array}{ccl}
h_{IIA}&=&2{\cal P}^- p^++\dot{X}^i\dot{X}^i+{X}^{i'}{X}^{i'}
+\hat{\mu}^2X^IX^I+4\hat{\mu}^2(X^2)^2
+4\hat{\mu}X^2X^1{}'
=0\\
t_{IIA}&=&p^+ X^{-'}+\dot{X}^i X^{i'}=0\\
\end{array}\right.\\
&&~~~{\cal P}^- =\dot{X}^-
 -4\mu^2 p^+
\left(X^IX^I+4(X^2)^2\right)-4\mu X^2{X}^1{}'\nn~~~.
\eea
The quantization is given by
\bea
X^1&=&w^\natural R\sigma+x^\natural+\alpha'p^\natural\tau
\nn\\&&+\frac{\sqrt{\alpha'}}{2\sqrt{2}}\displaystyle\sum_{n\neq 0}
\frac{1}{n}\left\{
(-1+\frac{\hat{\mu}}{n\rho_n})
(\alpha_ne^{-in\sigma}+\tilde{\alpha}_ne^{in\sigma})
e^{-i\hat{\mu}\tau}\right.\nn\\&&\quad\quad\quad~~~~~~~~~\left.
+(1+\frac{\hat{\mu}}{n\rho_n})
(\bar{\alpha}_ne^{-in\sigma}+\tilde{\bar{\alpha}}_ne^{in\sigma})
e^{i\hat{\mu}\tau}
\right\}e^{-in\rho_n\tau}\label{quantumX12}\\
X^2&=&-\frac{w^\natural R}{2\hat{\mu}}+
x^2\cos 2\hat{\mu}\tau
+\frac{\alpha'}{\hat{\mu}}p^2\sin 2\hat{\mu}\tau
\nn\\&&+\frac{\sqrt{\alpha'}}{2\sqrt{2}}\displaystyle\sum_{n\neq 0}
\frac{i}{n\rho_n}\left\{(\alpha_ne^{-in\sigma}-\tilde{\alpha}_ne^{in\sigma})
e^{-i\hat{\mu}\tau}
+(\bar{\alpha}_ne^{-in\sigma}-\tilde{\bar{\alpha}}_ne^{in\sigma})
e^{i\hat{\mu}\tau}
\right\}e^{-in\rho_n\tau}
\nn
\eea
with $\left[x^\natural,p^\natural\right]=i$
in addition to $X^I$ in the \bref{IIBXXX}.

The Noether charges in the IIA background are given as
\bea
\left\{\begin{array}{ccl}
Q_{[0]}^1&=&p^\natural\\
Q_{[0]}^{1*}&=&-p^+(x^\natural+Rw^\natural \pi)\\
Q_{[0]}^-&=&
-\frac{1}{2p^+}\left[
\displaystyle\sum_{i=2,3,\cdots,8}\left(\alpha'p^ip^i+\frac{\hat{\mu}^2}{\alpha'}x^ix^i\right)
+{\alpha'}(p^\natural)^2\right.\\&&\left.
+\displaystyle\sum_{n\neq 0}
\left\{
(\alpha_n^I\alpha_{-n}^I+\tilde{\alpha}_n^I\tilde{\alpha}_{-n}^I)
+\left(1+\frac{\hat{\mu}}{n\rho_n}\right)
\left(
\alpha_n\bar{\alpha}_{-n}+\tilde{{\alpha}}_n\tilde{\bar{\alpha}}_{-n}
\right)
\right\}
\right]
\end{array}
\right.\label{IIANoether}
\eea
and $Q_{[0]}^2,Q_{[0]}^{2*},Q_{[0]}^I,Q_{[0]}^{I*},Q_{[0]}^{IJ}$  are the same with the IIB case 
in \bref{IIBNoether}. 
Comparing the lightcone Hamiltonian in the type IIA background, $Q_{[0]}^-$, 
in \bref{IIANoether}
with the one in the type IIB background in \bref{IIBNoether} leads to
\bea
Q_{[0]}^-{}_{IIA}=Q_{[0]}^-{}_{IIB}~~\Leftrightarrow~~
\frac{1}{\alpha'}(wR)^2= \alpha'(p^\natural)^2~~~.
\eea
Then the zero-th order momentum charge $Q_{[0]}^1$ in the IIA background corresponds to the first order nonlocal momentum charge $Q_{[1]}^1$ in
the  IIB background in \bref{IIBnnc}
\bea
Q_{[0]}^1{}_{IIA}=-Q_{[1]}^1{}_{IIB}~~~.
\eea
In order to make this correspondence complete as
\bea
Q_{[0]}^{1*}{}_{IIA}=-Q_{[1]}^{1*}{}_{IIB}~~~,
\eea
let us add a 
Wess-Zumino term in the ``IIB" action
\bea
{\cal L}_{IIB;WZ}=\frac{1}{2}X^{\natural}\epsilon^{\alpha\beta}
 (\partial_\alpha J_\beta^1+\frac{4\mu}{2\pi\alpha'}\partial_\alpha X^2\partial_\beta X^+)~~~\label{WZBts}
 \eea
 where $X^\natural$ is a Lagrange multiplier to ensure
  the M.C. equation of $J^1$ in \bref{MCIIBS1}.
Under the $\fr{k}^{1*}=\frac{\partial}{\partial X^2}
+4\mu X^1\frac{\partial}{\partial X^-}
$ transformation 
the variation of ${\cal L}_{IIB;WZ}$ 
gives a new contribution to the current $J^{1*}_{IIB}$
as
\bea
\delta_{\lambda^{1*}}{\cal L}_{IIB;WZ}=\partial^\alpha\lambda^{1*}
\Delta J^{1*}_\alpha~~,~~
\Delta J^{1*}_\alpha=-\frac{1}{2\pi}X^\natural\epsilon_{\alpha\beta}\partial^\beta
X^+~~~.
\eea  
Then the first order nonlocal charge in the IIB background becomes
$Q_{[1]}^{1*}{}_{IIB}=p^+x^\natural$
where the zero mode of $X^\natural$ is $x^\natural$.
The commutator between $Q_{[1]}^{1}{}_{IIB}$ and 
$Q_{[1]}^{1*}{}_{IIB}$
is realized as same as 
the one of $Q_{[0]}^{1}{}_{IIA}$ and $Q_{[0]}^{1*}{}_{IIA}$.
The zero mode of $X^\natural$ is conjugate of $w$,
$\left[x^\natural,w R/\alpha' \right]=i$ with $x^1_+-x^1_-=x^\natural$ 
in the IIB background side.
 This WZ term \bref{WZBts} is turns out to be analogous to
 the Buscher T-duality transformation.

It is interesting that 
the M.C. equations for the one form current constructed from 
the IIA Noether currents \bref{S12JJ}
contain extra terms  
because of the IIA WZ term in \bref{actIIA}.
In order to compute the nonlocal currents in this case
this anomaly must be treated consistently.
We leave this problem for future investigation.

\par
\vskip 6mm
\section{Conclusions and discussion}

We have obtained nonlocal charges in terms of oscillators
in a flat background and the IIB and IIA pp-wave backgrounds.
For the flat background 
we have shown that 
the set of independent conserved nonlocal charges 
is the same before and after T-duality transformation
with interchanging odd and even-order charges;
for example the zero-th order charge in the T-dualized flat background
 coincides with
the ones in the original background.
T-duality interchanges
the momenta and the winding number
and the total Lorentz spin and
the relative Lorentz spin which is the difference between the left mover's spin and
 the  right mover's one.
Among an infinite set of nonlocal charges 
independent charges are the zero-th and first order charges which are 
the momenta and the winding number of zero mode
and the total spin and the relative spin
for zero mode and non-zero modes separately.

For the pp-wave background 
we have computed the nonlocal charges
and obtained expressions of the zero-th, first and 
second order ones 
in terms of oscillators.
Contrast to the flat case 
coefficients of the mode expansion in nonlocal charges 
are different, so
there exist an infinite set of independent nonlocal charges
for the pp-wave case.
We have  shown that
the zero-th order charges in the T-dualized  pp-wave background, 
 the IIA pp-wave background, 
are included as a part of  
an infinite set of  nonlocal charges in 
 the original IIB pp-wave background.
Since we perform the lightcone quantization
the lightcone Hamiltonians for the type IIB and the type IIA backgrounds
are equal by  T-duality.
This equality leads to identification of the 
modes in both sides. 
As a result the zero-th order momentum charge in the IIA pp-wave side,
 $Q_{[0]}^{1}{}_{IIA}$,
corresponds to the first order nonlocal momentum charge
in the IIB pp-wave side, $Q_{[1]}^{1}{}_{IIB}$.
In order to make this correspondence complete
the zero-th order charge in the IIA pp-wave side, $Q_{[0]}^{1*}{}_{IIA}$,
should correspond to
 the non-zero value of the first order charge $Q_{[1]}^{1*}{}_{IIB}$ in the IIB pp-wave side. 
Then we introduce a WZ term for the ``IIB" pp-wave 
background 
in such a way that this term causes non-zero 
value of $Q_{[1]}^{1*}{}_{IIB}$
satisfying the corresponding algebra.
It turns out that the Lagrange multiplier of the WZ term
is a variable conjugate
to the winding mode. 
This term is nothing but the term used in the
 Buscher T-duality transformation.
In another word one can introduce the conjugate coordinate
to the winding mode by adding the
WZ term  $\grave{\rm a}$ la  Buscher's T-duality transformation.
The completeness of this correspondence
requires the IIB side to add the 
$B_{\mu\nu}$ field as a target space 
interpretation of the WZ term
and to include 
the relative coordinate $x^\natural=x_+-x_-$,
so these dual degrees of freedom are hidden in also 
the ``IIB" side.
Then it is natural to formulate string theories by
``two-vierbein formalism"  
\cite{Siegel:1993xq}
and it may be generalize to the general field theories.
It may be interesting to relate the issue to 
the finite size effect of the integrable system
\cite{Mann:2004jr}.

In this paper we clarified the procedure of
constructing the nonlocal charges for 
inhomogeneous SO($n$) cosets such as
a flat and the pp-wave background cases;
the basic currents which satisfy the conservation law 
are set to be the Noether currents of the action,
and the  ``flatness" condition 
is examined.
Based on these currents nonlocal charges are constructed inductively.
The conserved charges are obtained by the integration 
along the boundary of the semi-infinite strip 
cut open the cylinderical worldsheet.
We could not compute the higher order nonlocal charges
in the type IIA pp-wave background,
since the ``flatness" condition of
the IIA Noether currents
includes extra terms
caused by the WZ term.
We know that the WZ term  produces
a topological center in the Noether charge algebra,
where charges can be constructed but
the ground state is only  invariant under a part of the symmetries.
If this IIA theory and the IIB  theory are really T-dual,
then the infinite set of nonlocal charges also
exist in both theories. 
So  
there may be a further generalization of the procedure
constructing nonlocal charges 
for ``non-flat" systems.

Nonlocal charges carry T-dual information as we have shown.
T-duality in the pp-wave space may trace back to the
one in the AdS space
where an infinite set of nonlocal charges   
exist.
So application of our analysis to the AdS space
may be  possible in the classical level.
It is curious how 
 nonlocal charges in
the holographic dual theories 
realize T-dual information.
Generalization involving U-duality may be 
interesting, and we leave these problem for future investigation.

\par\vskip 6mm

\noindent{\bf Acknowledgments}

The authors would like to thank F.A. Bais, R.Roiban,
W. Siegel,  B.C. Vallilo, K. Yoshida,
 and E. Witten for fruitful discussion.
M.H. would like to thank  the
 2005 Simons workshop on Mathematics and Physics
 for warm hospitality  and providing a stimulating environment
 where part of this work was done.
S.M. was supported by Grant-in-Aid
for Scientific Research (C)(2)\#16540273 from
The Ministry of Education, Culture, Sports, Science
and Technology.

\par\vskip 6mm


\begin{thebibliography}{99}
\bibliographystyle{unsrt}
%
\setlength{\itemsep}{0.0in}

  
\bibitem{BMN}
D.~Berenstein, J.~M.~Maldacena and H.~Nastase,
``Strings in flat space and pp waves from N = 4 super Yang Mills,''
JHEP {\bf 0204} (2002) 013
[arXiv:hep-th/0202021].

\bibitem{GKP2}
S.~S.~Gubser, I.~R.~Klebanov and A.~M.~Polyakov,
``A semi-classical limit of the gauge/string correspondence,''
Nucl.\ Phys.\ B {\bf 636} (2002) 99
[arXiv:hep-th/0204051].


\bibitem{FT}
S.~Frolov and A.~A.~Tseytlin,
``Semiclassical quantization of rotating superstring in AdS$_5\times$ S$^5$,''
JHEP {\bf 0206} (2002) 007
[arXiv:hep-th/0204226].

  
\bibitem{MZ}
J.~A.~Minahan and K.~Zarembo,
``The Bethe-ansatz for N = 4 super Yang-Mills,''
JHEP {\bf 0303} (2003) 013
[arXiv:hep-th/0212208]. 
  
  
  \bibitem{Frolov:2003xy}
  S.~Frolov and A.~A.~Tseytlin,
  ``Rotating string solutions: AdS/CFT duality in non-supersymmetric
  sectors,''
  Phys.\ Lett.\ B {\bf 570} (2003) 96
  [arXiv:hep-th/0306143].
  
\bibitem{Beisert:2003tq}
  N.~Beisert, C.~Kristjansen and M.~Staudacher,
  ``The dilatation operator of N = 4 super Yang-Mills theory,''
  Nucl.\ Phys.\ B {\bf 664} (2003) 131
  [arXiv:hep-th/0303060];\\
N.~Beisert,
``The complete one-loop dilatation operator of N = 4 super Yang-Mills theory,''
Nucl.\ Phys.\ B {\bf 676} (2004) 3
[arXiv:hep-th/0307015].


\bibitem{Arutyunov:2003uj}
  G.~Arutyunov, S.~Frolov, J.~Russo and A.~A.~Tseytlin,
  ``Spinning strings in AdS(5) x S**5 and integrable systems,''
  Nucl.\ Phys.\ B {\bf 671} (2003) 3
  [arXiv:hep-th/0307191];
  G.~Arutyunov, J.~Russo and A.~A.~Tseytlin,
  ``Spinning strings in AdS(5) x S**5: New integrable system relations,''
  Phys.\ Rev.\ D {\bf 69} (2004) 086009
  [arXiv:hep-th/0311004].

  \bibitem{Beisert:2003yb}
  N.~Beisert and M.~Staudacher,
  ``The N = 4 SYM integrable super spin chain,''
  Nucl.\ Phys.\ B {\bf 670} (2003) 439
  [arXiv:hep-th/0307042].
  
  \bibitem{Zarembo:2004hp}
   N.~Beisert,
  ``The dilatation operator of N = 4 super Yang-Mills theory and
  integrability,''
  Phys.\ Rept.\  {\bf 405} (2005) 1
  [arXiv:hep-th/0407277];\\
  K.~Zarembo,
  ``Semiclassical Bethe ansatz and AdS/CFT,''
  Comptes Rendus Physique {\bf 5} (2004) 1081
  [Fortsch.\ Phys.\  {\bf 53} (2005) 647]
  [arXiv:hep-th/0411191];\\
  J.~Plefka,
 ``Spinning strings and integrable spin chains in the AdS/CFT
 correspondence,''
  arXiv:hep-th/0507136.
 
 
  \bibitem{LP}
M.~L$\ddot{\rm u}$scher, 
``Quantum nonlocal charges and absence of particle production in the 
two-dimensional nonlinear sigma model,''
Nucl.\ Phys.\ B {\bf 135} (1978) 1, \\ 
M.~L$\ddot{\rm u}$scher and K.~Pohlmeyer,
 ``Scattering of massless lumps and nonlocal charges in the
	two-dimensional classical nonlinear sigma model,''
Nucl.\ Phys.\ B {\bf 137} (1978) 46. 
 \bibitem{Lipatov}
L.N. Lipatov, 
`High-energy asymptotics of multicolor QCD and exactly solvable lattice
  models,''
 JETP Lett. {\bf 59} (1994) 596  [arXiv:hep-th/9311037];\\
G. Korchemsky and L. Faddeev, 
`High-energy QCD as a completely integrable model,''
 Phys. Lett.  B{\bf 342} (1995) 311  [arXiv:hep-th/9404173].



\bibitem{BIZZ}
E.~Brezin, C.~Itzykson, J.~Zinn-Justin and J.~B.~Zuber,
``Remarks about the existence of nonlocal charges in two-dimensional models,''
Phys.\ Lett.\ B {\bf 82} (1979) 442. 



\bibitem{Buscher:1987sk}
  T.~H.~Buscher,
 ``A Symmetry Of The String Background Field Equations,''
  Phys.\ Lett.\ B {\bf 194} (1987) 59;
 ``Path Integral Derivation Of Quantum Duality In Nonlinear Sigma Models,''
  Phys.\ Lett.\ B {\bf 201} (1988) 466.
     \bibitem{BPR}
I.~Bena, J.~Polchinski and R.~Roiban,
``Hidden symmetries of the AdS$_5 \times$ S$^5$ superstring,''
Phys.\ Rev.\ D {\bf 69} (2004) 046002
[arXiv:hep-th/0305116].

   \bibitem{Mandal:2002fs}
  G.~Mandal, N.~V.~Suryanarayana and S.~R.~Wadia,
  ``Aspects of semiclassical strings in AdS(5),''
  Phys.\ Lett.\ B {\bf 543} (2002) 81
  [arXiv:hep-th/0206103].

\bibitem{HY} 
 M.~Hatsuda and K.~Yoshida,
  ``Classical integrability and super Yangian of superstring on AdS(5) x
  S**5,'' 
  Adv. Theor. Math. Phys. 9 (2005) 703,
  [arXiv:hep-th/0407044].

\bibitem{Vallilo:2003nx}
  B.~C.~Vallilo,
  ``Flat currents in the classical AdS(5) x S**5 pure spinor superstring,''
  JHEP {\bf 0403} (2004) 037
  [arXiv:hep-th/0307018].
\bibitem{Berkovits:2004jw}
  N.~Berkovits,
  ``BRST cohomology and nonlocal conserved charges,''
  JHEP {\bf 0502} (2005) 060
  [arXiv:hep-th/0409159];
  ``Quantum consistency of the superstring in AdS(5) x S**5 background,''
  JHEP {\bf 0503} (2005) 041
  [arXiv:hep-th/0411170].


\bibitem{Hou:2004ru}
  B.~Y.~Hou, D.~T.~Peng, C.~H.~Xiong and R.~H.~Yue,
``The affine hidden symmetry and integrability of type IIB superstring in
AdS(5) x S**5,''
  arXiv:hep-th/0406239.
 
  \bibitem{Wolf:2004hp}
  M.~Wolf,
``On hidden symmetries of a super gauge theory and twistor string theory,''
  JHEP {\bf 0502} (2005) 018
  [arXiv:hep-th/0412163].


\bibitem{Das:2004hy}
  A.~Das, J.~Maharana, A.~Melikyan and M.~Sato,
  ``The algebra of transition matrices for the AdS(5) x S**5 superstring,''
  JHEP {\bf 0412} (2004) 055
  [arXiv:hep-th/0411200].

\bibitem{Alday:2005gi}
  L.~F.~Alday, G.~Arutyunov and A.~A.~Tseytlin,
  ``On integrability of classical superstrings in AdS(5) x S**5,''
  JHEP {\bf 0507} (2005) 002
  [arXiv:hep-th/0502240].


  \bibitem{Young:2005jv}
  C.~A.~S.~Young,
  ``Non-local charges, Z(m) gradings and coset space actions,''
  Phys.\ Lett.\ B {\bf 632} (2006) 559
  [arXiv:hep-th/0503008].

\bibitem{Miller:2006bu}
  B.~H.~Miller,
  ``Conserved charges in the principal chiral model on a supergroup,''
  arXiv:hep-th/0602006.
  

\bibitem{Hatsuda:2005te}
  M.~Hatsuda,
  ``Sugawara form for AdS superstring,''
  Nucl.\ Phys.\ B {\bf 730} (2005) 364
  [arXiv:hep-th/0507047].
  

\bibitem{Alday:2003zb}
  L.~F.~Alday,
``Non-local charges on AdS(5) x S**5 and pp-waves,''
  JHEP {\bf 0312} (2003) 033
  [arXiv:hep-th/0310146].

\bibitem{Michelson:2002wa}
  J.~Michelson,
  ``(Twisted) toroidal compactification of pp-waves,''
  Phys.\ Rev.\ D {\bf 66} (2002) 066002
  [arXiv:hep-th/0203140].
  \bibitem{Alishahiha:2002nf}
  M.~Alishahiha, M.~A.~Ganjali, A.~Ghodsi and S.~Parvizi,
  ``On type IIA string theory on the PP-wave background,''
  Nucl.\ Phys.\ B {\bf 661} (2003) 174
  [arXiv:hep-th/0207037].
  
\bibitem{Mizoguchi:2003be}
  S.~Mizoguchi, T.~Mogami and Y.~Satoh,
  ``A note on T-duality of strings in plane-wave backgrounds,''
  Phys.\ Lett.\ B {\bf 564} (2003) 132
  [arXiv:hep-th/0302020].
  
    
  \bibitem{Berkovits:1999zq}
  N.~Berkovits, M.~Bershadsky, T.~Hauer, S.~Zhukov and B.~Zwiebach,
  ``Superstring theory on AdS(2) x S(2) as a coset supermanifold,''
  Nucl.\ Phys.\ B {\bf 567} (2000) 61
  [arXiv:hep-th/9907200].
  
  \bibitem{Frolov:2006cc}
  S.~Frolov, J.~Plefka and M.~Zamaklar,
  ``The AdS5xS5 Superstring in Light-Cone Gauge and its Bethe Equations,''
  arXiv:hep-th/0603008.
\bibitem{Siegel:1993xq}
  W.~Siegel,
  ``Two vierbein formalism for string inspired axionic gravity,''
  Phys.\ Rev.\ D {\bf 47} (1993) 5453
  [arXiv:hep-th/9302036].

\bibitem{Mann:2004jr}
  N.~Mann and J.~Polchinski,
  ``Finite density states in integrable conformal field theories,''
  arXiv:hep-th/0408162.
  
  
  
 \end{thebibliography}
\end{document}